\documentclass[aps,prl,reprint,superscriptaddress,nofootinbib]{revtex4-2}
\usepackage{graphicx}
\usepackage[english]{babel}
\usepackage{braket}
\usepackage{float}
\usepackage{amsmath}
\usepackage{xcolor,soul} 

\newcommand\Ca{$^{40}\text{Ca}^+$}
\newcommand\PenningN{12}
\newcommand\ratioPenning{0.599 488 813 3(2)}
\newcommand\ratioRF{0.599 488 813(6)}
\newcommand\PenningFrac{0.34~ppb}

\begin{document}


\title{Dual-Platform Precision Measurement of the $3^2D_{5/2}$ to $4^2S_{1/2}$ $g$-Factor Ratio in \Ca}



\author{Brian J. McMahon}
\thanks{These authors contributed equally to this work}
\author{Vikram S. Sandhu}
\thanks{These authors contributed equally to this work}
\author{John M. Gray}
\affiliation{Georgia Tech Research Institute, Atlanta, Georgia 30332, USA}
\author{Creston D. Herold}
\affiliation{Georgia Tech Research Institute, Atlanta, Georgia 30332, USA}
\author{Christopher M. Seck}
\author{Gilles Buchs}
\affiliation{Quantum Information Science Section, Computational Sciences and Engineering Division, Oak Ridge National Laboratory, Oak Ridge, Tennessee, 37381, USA}
\author{Kenton R. Brown}
\affiliation{Georgia Tech Research Institute, Atlanta, Georgia 30332, USA}
\author{Brian C. Sawyer}
\email[]{Corresponding author Brian.Sawyer@gtri.gatech.edu}
\affiliation{Georgia Tech Research Institute, Atlanta, Georgia 30332, USA}

\begin{abstract}
We report precision measurements of the ratio of Land\'e $g$ factors between the $3^2D_{5/2}$ and $4^2S_{1/2}$ states of a single trapped \Ca~ion. The measurements are performed in two distinct ion trap apparatus: a cryogenic surface electrode radiofrequency Paul trap and a room-temperature permanent magnet Penning trap. The Penning trap measurements rely on resonant microwave excitation of magnetic sublevels and yield a ratio of \ratioPenning, which is a more than 40-fold uncertainty reduction compared to previous work. The radiofrequency trap measurements utilize optical electric quadrupole transitions and yield a concurring value of \ratioRF. We estimate that systematic shifts for each system are well below the respective statistical uncertainty. 
\end{abstract}

\maketitle

\textit{Introduction.} Precision measurements of atomic and subatomic properties play an important role in advancing our understanding of fundamental physics and in refining high-accuracy theoretical models of atomic structure. Trapped atomic ions are among the most versatile and robust platforms for furthering both applied quantum technologies and fundamental physics understanding. Their exceptional degree of isolation from external noise, long transition coherence times, and high-fidelity quantum state preparation and control have made trapped-ion systems cornerstones in fields ranging from precision metrology~\cite{tommaseo_mathsfg_scriptscriptstyle_2003,blaum_high-accuracy_2006,shiga_diamagnetic_2011,arnold_prospects_2015,pezze_quantum_2018,brewer__2019,zhiqiang_176_2023,arnold_optical_2025,filzinger_multi-ion_2026} to quantum information science~\cite{britton_engineered_2012,monroe_programmable_2021,noel_measurement-induced_2022,moses_race-track_2023,jain_penning_2024,loschnauer_scalable_2025}. {\let\thefootnote\relax\footnotetext{This manuscript has been authored in part by UT-Battelle, LLC, under contract DE-AC05-00OR22725 with the US Department of Energy (DOE). The US government retains and the publisher, by accepting the work for publication, acknowledges that the US government retains a non-exclusive, paid-up, irrevocable, world-wide license to publish or reproduce the submitted manuscript version of this work, or allow others to do so, for US government purposes. DOE will provide public access to these results of federally sponsored research in accordance with the DOE Public Access Plan (https://energy.gov/doe-public-access-plan).}}

In this Letter, we present a dual-platform investigation of the ratio of Land\'e $g$-factors for the metastable $3^2D_{5/2}$ and ground $4^2 S_{1/2}$ states of a single trapped \Ca~ion, measured using two complementary experimental setups: a cryogenic surface electrode radiofrequency (rf) Paul trap \cite{hartsell_design_2026} and a compact room-temperature Penning trap \cite{mcmahon_individual-ion_2024,mcmahon_efficient_2026}. By achieving a measurement precision below one part per billion in the Penning trap system and corroborating the results with the rf trap system, we reduce the uncertainty of this $g$-factor ratio by $>40\times$~\cite{ma_precision_2024,zhang_liquid-nitrogen-cooled_2026} and approximately three orders of magnitude~\cite{chwalla_absolute_2009,guevara-bertsch_absolute_2026} compared to previous studies. Our work not only enhances the accuracy of \Ca~as a model system for quantum information, timing, and precision metrology but also demonstrates the utility of combining distinct trapping technologies for high-precision spectroscopic measurements. The $g$-factors of bound states of electrons are sensitive to multi-electron interactions, quantum electrodynamics, and nuclear corrections~\cite{tommaseo_mathsfg_scriptscriptstyle_2003,beverini_g_1998, vogel_trap-assisted_2010,sahoo_relativistic_2017,kumar_diverse_2018}. Precision measurements of bound-electron $g$-factors and their ratios therefore provide for stringent tests of atomic structure models.

\begin{figure}
	\includegraphics[scale=0.42]{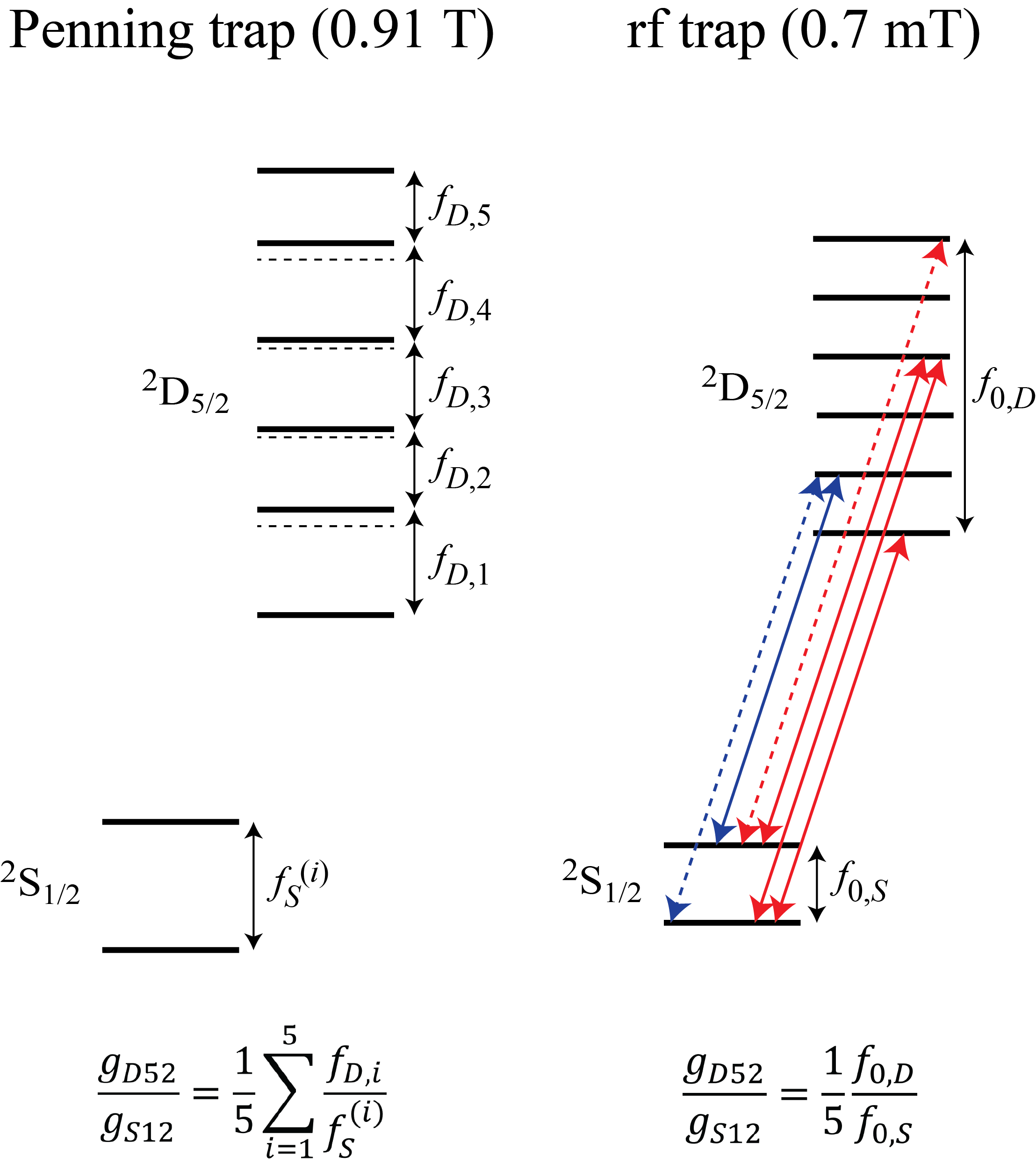}
    \caption{\label{fig:levels} Illustration of the two spectroscopy techniques used for the dual-platform $g$-factor ratio measurement. (Left) In the 0.91~T magnetic field of the compact Penning trap, the $\left( D_{5/2},~|m_J|\leq3/2\right)$ sublevels experience strong nonlinear Zeeman shifts due to coupling of the $D_{5/2}$ and $D_{3/2}$ manifolds. Solid lines reflect the actual sublevel Zeeman shifts relative to the first-order approximation given by the dashed lines. (Right) In the 0.7~mT field of the rf trap, Ramsey experiments are used to measure the transition frequencies $f_{0,S}$ and $f_{0,D}$. Sequences of 729~nm laser pulses, applied sequentially from left to right, are used to prepare the desired superposition states within the $S_{1/2}$ (blue) or $D_{5/2}$ (red) manifolds. Dashed arrows indicate $\pi/2$ pulses, while solid arrows indicate $\pi$ pulses.}
\end{figure}

There exists disagreement between recent measurements of the $g$-factor ratio between the $3^2D_{5/2}$ and $4^2S_{1/2}$ states of \Ca~\cite{chwalla_absolute_2009,ma_precision_2024,zhang_liquid-nitrogen-cooled_2026,guevara-bertsch_absolute_2026}. The reported values are $0.599~490~58(15)$~\cite{chwalla_absolute_2009}, $0.599~488~79(2)$~\cite{ma_precision_2024}, $0.599~488~818(9)$~\cite{zhang_liquid-nitrogen-cooled_2026}, and $0.599~488~1(5)$~\cite{guevara-bertsch_absolute_2026}. The results of Ref.~\cite{chwalla_absolute_2009} and Refs.~\cite{ma_precision_2024,zhang_liquid-nitrogen-cooled_2026,guevara-bertsch_absolute_2026} are offset from one another by more than ten combined standard deviations. Theoretical estimates of this ratio are currently limited to $\sim30$~ppm uncertainty due to large electron correlation contributions~\cite{kumar_diverse_2018}, and that uncertainty is much larger than those of Refs.~\cite{chwalla_absolute_2009,ma_precision_2024,zhang_liquid-nitrogen-cooled_2026,guevara-bertsch_absolute_2026}. Recent work highlights the potential for non-negligible AC Zeeman shifts in precision measurements using surface electrode rf ion traps~\cite{gan_oscillating-magnetic-field_2018,arnold_precision_2020,ivory_ac_2024,guevara-bertsch_absolute_2026}, which suggests that multi-platform (i.e. rf and Penning trap) comparisons may be advantageous.

For both systems described in this Letter, we compare the `full span' $\left( m_J=\pm5/2 \right)$ transition frequency of the $D_{5/2}$ manifold to that of the $\left( m_J=\pm1/2 \right)$ $S_{1/2}$, where $m_J$ is the magnetic sublevel quantum number (see Fig.~\ref{fig:levels}). This measurement approach removes a number of systematic shifts in both the high- and low-magnetic-field systems. To lowest order, the frequency shifts of the $m_J$ sublevels in the presence of an external magnetic field $B$ are $-g_J \mu_b B m_J / h$, where $h$ is Planck's constant, $\mu_b$ is the Bohr magneton, and $g_J$ is the Land\'e $g$-factor of state $J$. Our measurement of the full span of $D_{5/2}$ sublevels avoids nonlinear Zeeman, diamagnetic~\cite{beverini_g_1998}, and electric quadrupole frequency shifts.

\textit{Experiment (Penning trap).} The compact Penning trap is described in recent publications~\cite{mcmahon_individual-ion_2024, mcmahon_efficient_2026}. Briefly, the trap assembly consists of two composite, radially-magnetized SmCo ring magnets producing a uniform axial magnetic field of $\sim0.9145~\text{T}$ near the trap geometric center. The additional electrostatic fields required for ion confinement are generated from two identical printed circuit boards (PCBs) mounted between the two ring magnets. For this work, we confine a single \Ca~ion produced via photoionization of a Ca atomic beam with trap eigenfrequencies of $f_z=200$~kHz, $f_+=280$~kHz, and $f_-=72$~kHz. Here $f_z$, $f_+$, and $f_-$ denote the axial, modified cyclotron, and magnetron motional frequencies, respectively.

The single \Ca~ion is Doppler cooled using laser excitation near 397~nm ($S_{1/2}\leftrightarrow P_{1/2}$), 866~nm ($D_{3/2}\leftrightarrow P_{1/2}$), and 854~nm ($D_{5/2}\leftrightarrow P_{3/2}$). We apply two different laser frequencies near 397~nm separated by $\sim34$~GHz, four optical `repump' tones near 866~nm generated using a broadband electro-optic modulator to clear the $D_{3/2}$ population leakage, and four tones near 854~nm~\cite{koo_doppler_2004,mcmahon_doppler-cooled_2020}. The repump frequencies near 854~nm are required due to the weak $P_{1/2} \rightarrow D_{5/2}$ spontaneous decay channel induced by magnetic coupling of the $P_{1/2}$ and $P_{3/2}$ sublevels~\cite{crick_magnetically_2010}.

We perform axial sub-Doppler cooling of \Ca~ using two-photon dark resonance cooling (DRC) with frequency-stabilized 397~nm and 866~nm laser beams followed by resolved-sideband cooling on the $S_{1/2}\leftrightarrow D_{5/2}$ electric quadrupole transition at 729~nm~\cite{mcmahon_efficient_2026}. We typically achieve an axial mode occupation of $\bar{n}_z < 1$ in 2~ms. Following axial sub-Doppler cooling, we perform a parametric exchange of motional occupation between the cold axial mode and one of the radial modes as in~\cite{cornell_mode_1990} and repeat the axial sub-Doppler cooling routine. After performing the parametric exchange and sub-Doppler cooling for both the modified cyclotron and magnetron modes, we typically achieve final simultaneous mode occupations of $\bar{n}_z = 0.4(2)$, $\bar{n}_+ = 12(2)$, and $\bar{n}_- = 20(3)$ with a total sub-Doppler cooling time of 6~ms~\cite{mcmahon_efficient_2026}.

\begin{figure}
	\includegraphics[scale = 0.31]{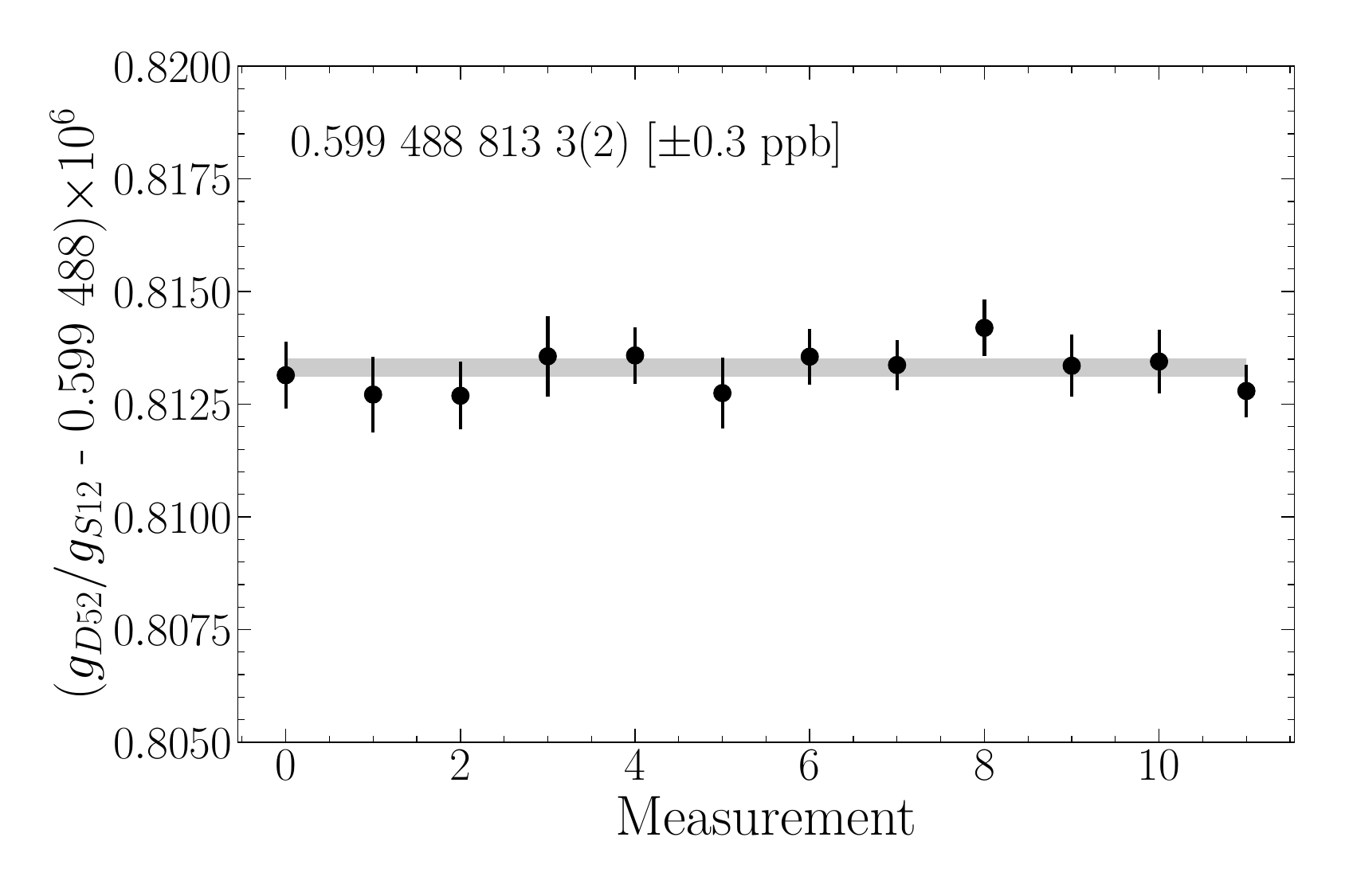}
	\caption{\label{fig:PenningData} Repeated $g$-factor ratio measurements in the compact Penning trap with corresponding standard errors. The light gray region corresponds to the weighted standard error bounds computed from all measurements. We report a $g$-factor ratio of \ratioPenning, corresponding to a fractional statistical uncertainty of~\PenningFrac.}
\end{figure}

Following sub-Doppler cooling, we perform a series of optical state preparation and resonant microwave pulses to measure transition frequencies within the $D_{5/2}$ and $S_{1/2}$ state manifolds (see Fig.~\ref{fig:levels}, left). We first apply a single 397~nm laser beam along with all 866~nm repump tones to efficiently prepare the $|S_{1/2},m_J=-1/2\rangle$ state. For ground state measurements, we then perform a Rabi interrogation of the $|S_{1/2},-1/2\rangle \leftrightarrow |S_{1/2},+1/2\rangle$ transition by applying a near-resonant $\pi$ pulse at $\sim26.5~\text{GHz}$ for 3~ms using a microwave horn mounted outside the vacuum chamber. Any remaining $|S_{1/2},-1/2\rangle$ population is then coherently transferred (i.e. shelved) to the $D_{5/2}$ state using a frequency-stabilized 729~nm laser beam, and the total $S_{1/2}$ population is finally detected for 3~ms using the 397~nm and 866~nm laser beams. The 3~ms Rabi excitation yields a minimum ground state resonance full-width-at-half-maximum (FWHM) of $\sim270$~Hz, which allows for a determination of the center frequency to $\sim10$~Hz $(\pm0.4~\text{ppb})$ in a frequency scan.

The measurement procedure for the $D_{5/2}$ manifold is more involved than for the $S_{1/2}$ due to its six $m_J$ sublevels. We have opted to remove systematic shifts that affect the inner $\left( |m_J|\leq3/2 \right)$ states by measuring the `full span' frequency separation of the $m_J=\pm5/2$ levels. As with the ground state, the metastable transition measurement sequence begins with initial state preparation to $|S_{1/2},-1/2\rangle$. This is followed by a transfer of the full $|S_{1/2},-1/2\rangle$ population to the $|D_{5/2}, -3/2\rangle$ using the resonant 729~nm laser beam. We then measure one of the five $|\Delta m_J|=1$ transition frequencies using Rabi spectroscopy. The Rabi $\pi$ pulses are induced from a second horn emitting resonant microwave radiation near 15.4~GHz.

The large magnetic field of the Penning trap induces nonlinear Zeeman shifts of the $\left( |m_J|\leq3/2 \right)$ levels due to coupling of the $D_{5/2}$ and $D_{3/2}$ sublevels, shifting adjacent microwave transitions by $\sim10$~MHz and allowing us to transfer population coherently throughout the $D_{5/2}$ manifold using a series of resonant $\pi$ pulses. For each metastable transition frequency measurement, we first transfer population from the $|D_{5/2}, -3/2\rangle$ shelving state to the state of interest using a sequence of distinct resonant $\pi$ pulses near 15.4~GHz each with a duration of $30 - 40~\mu\text{s}$. We then apply a variable-frequency Rabi excitation pulse to the target metastable transition for a 2.7~ms duration. Finally, we invert the initial sequence of resonant $\pi$ pulses to transfer population back to $m_J=-3/2$ before applying an optical de-shelving $\pi$ pulse at 729~nm. A measurement of the $S_{1/2}$ population then provides the relative metastable sublevel populations for the given Rabi frequency scan.

Magnetic field stability is a primary concern when measuring \textit{g}-factors. Our compact permanent magnet Penning trap exhibits magnetic field drift both from external laboratory fields and from internal temperature shifts of the SmCo magnets themselves. The SmCo magnets have a specified fractional temperature coefficient of $-1\times10^{-5} / K$. For the measurements of Fig.~\ref{fig:PenningData}, we observe a ground state transition frequency drift of $3.95(1)$~Hz/s, indicating a magnetic field drift rate of $140.8(4)$~pT/s. This is consistent with a magnet temperature drift of $\sim15~\mu$K/s. While very slow magnetic field drift will not shift our reported $g$-factor ratio, field drift \textit{during} the measurement of the relevant $D_{5/2}$ and $S_{1/2}$ microwave transitions could systematically shift the final computed ratio. To mitigate this, we measure the $S_{1/2}$ transition frequency in succession with each of the five $D_{5/2}$ sublevel transition measurements, alternating between $S_{1/2}$ and $D_{5/2}$ measurements. The five $D_{5/2}$ to $S_{1/2}$ transition frequency ratios are then averaged to obtain one $g$-factor ratio $(g_{D52} / g_{S12})$ value as: 

\begin{equation}
\frac{g_{D52}}{g_{S12}} = \frac{1}{5}\sum_{i=1}^5 \frac{f_{D,i}}{f_S^{(i)}},
\end{equation}
where $f_S^{(i)}$ is the ground state transition frequency measured immediately before the $D_{5/2}$ transition frequency $f_{D,i}$. Recording the individual transition ratios in this way removes the drift of the linear Zeeman shift between measurements of the five $D_{5/2}$ transitions due to the trap magnetic field variation. We choose not to phase-synchronize these experiments with the $60$~Hz laboratory power since we observe a systematic shift of microwave transition frequencies that is phase coherent with the laboratory-power frequency. With a measurement duration of 75~s per frequency ratio, we estimate that the remaining second-order Zeeman shift contribution to the fractional error in $(g_{D52}/g_{S12})$ due to our $140.8(4)$~pT/s magnetic field drift is a negligible $7\times10^{-13}$ (see Table~\ref{table:error_budget}). 

The \PenningN~sets of transition frequency measurements shown in Fig.~\ref{fig:PenningData} give a final ratio of \ratioPenning, corresponding to a weighted fractional standard error of $\pm$\PenningFrac~limited by statistics. We quantify systematic shifts for this measurement scheme in Table~\ref{table:error_budget}. We identify three leading systematic shifts for the Penning trap measurements of the $m_J=\pm5/2$ frequency difference: AC Stark shifts due to leakage of 397~nm and 866~nm laser light during Rabi excitations when the acousto-optic modulators are switched off, AC Zeeman shifts due to near-resonance excitation of neighboring `spectator' transitions within the $D_{5/2}$ manifold, and linear magnetic field drift \emph{between} consecutive measurements of a $D_{5/2}$ and $S_{1/2}$ transition. All estimated systematic shifts in Table~\ref{table:error_budget} are well below the statistical uncertainty, so we do not apply systematic corrections to the measured ratio.

\textit{Experiment (rf trap).} The surface-electrode rf trap is housed inside a cryogenic vacuum chamber described in ~\cite{hartsell_design_2026}. Notably, the chamber features in-vacuum magnetic shielding layers at both room temperature and cryogenic temperatures. The quantization axis is set by an array of in-vacuum permanent magnets producing a 0.7~mT field at the ion location. Ramsey coherence measurements on the ground state $S_{1/2} (m_J=-1/2\leftrightarrow +1/2)$ qubit with 28~GHz/T field sensitivity yield a coherence time of 23(1)~ms without dynamical decoupling.

\begin{table*}
\centering
\begin{tabular}{|l | c | c|} 
\hline
$g$ Factor Ratio Error Source & Penning Trap $(\times10^{-10})$ & rf Trap $(\times10^{-10})$ \\ 
\hline\hline
\parbox[t]{2in}{\raggedright Statistical Uncertainty} & $3.4$ & $97.2$ \\
AC Stark shift (light leakage) & $<1.1$ & $3.8$  \\
AC Zeeman shift (off-resonant) & $<0.0014$ & - \\
Linear magnetic field drift & $0.007$ & - \\
\hline
\end{tabular}
\caption{Table of fractional statistical and systematic error estimates for both ion trap systems. We have applied no systematic corrections to our reported $g$-factor ratios since all identified systematic shifts in a given system are smaller than the respective statistical error.}
\label{table:error_budget}
\end{table*}

A single \Ca~ion is confined 34~$\mu$m above the trap surface using a trap rf drive frequency of $144$~MHz. The ion's axial secular frequency is $\sim2$~MHz and its radial secular frequencies are $\sim8$~MHz. The three secular modes are Doppler cooled using detuned 397~nm light with additional 866~nm light for repumping. Resolved sideband cooling via the $S_{1/2}\leftrightarrow D_{5/2}$ transition at 729~nm combined with 854~nm deshelving and 866~nm repumping reduces the axial mean phonon number to $\bar{n}_z \approx 0.1$.

Because neighboring $\Delta m_J=1$ transitions within the $D_{5/2}$ manifold are nearly degenerate at low magnetic field, the Rabi-style measurements performed at higher magnetic field in the Penning trap cannot be used in the rf trap. We choose instead to perform a Ramsey-style measurement, creating the required superposition of magnetic sublevels using a series of pulses at 729~nm between $S_{1/2}$ and $D_{5/2}$, as illustrated in Fig.~\ref{fig:levels} (right). This strategy confers the additional benefit of eliminating unwanted AC Stark shifts during the measurement. To measure a given qubit transition, the ion is first initialized in one of the Zeeman sublevels of the $S_{1/2}$ manifold. We then apply a preparation pulse sequence using 729~nm light to prepare an equal superposition of the desired states. For example, to prepare a Ramsey superposition of the $S_{1/2} (m_J=-1/2)$ and $S_{1/2} (m_J=+1/2)$ states, we apply a $\pi/2$ pulse on the $S_{1/2} (m_J=-1/2) \leftrightarrow D_{5/2} (m_J=-3/2)$ transition followed by a $\pi$ pulse between $D_{5/2} (m_J=-3/2)$ and $S_{1/2} (m_J=+1/2)$. The superposition undergoes free evolution at frequency $f_0$ during a Ramsey delay $\tau$. We then apply an analysis pulse sequence: the inverse of the preparation sequence.

The final 729~nm $\pi /2$ pulse in the analysis sequence is applied with a phase $\phi = \phi_{est} - 2\pi (f_{est}+\delta f) \tau$ which dictates the final population in the $D_{5/2}$ manifold. Here, $f_{est}$ is an estimate of the desired transition frequency based on previous calibrations, and $\delta f$ represents an offset to this frequency which we vary to determine the true resonance (where the final $S_{1/2}$ population is minimized). The phase offset $\phi_{est}$ represents the estimated phase difference between preparation and analysis pulse sequences. Because this phase is not well calibrated a priori, we treat the true phase offset, $\phi_0$, as a second unknown parameter and extract both $\phi_0$ and $f_0$, the true transition frequency, from our measurements as described below. Following all 729~nm pulses, the population in $S_{1/2}$ is determined via state-dependent fluorescence.

Varying $\delta f$ (and nothing else) at two different delays $\tau_1$ and $\tau_2$ yields two distinct values $\delta f_1$ and $\delta f_2$ which minimize the final measured population. Note that, if $\phi_{est}$ were well calibrated, these two values would be equal, but this is not the case. However, these two values can be used to estimate the true transition frequency, $f_0$, because $\delta f_1$ and $\delta f_2$ are related to the actual and estimated transition frequencies as
\begin{align*}
    \delta f_1&=f_0-f_{est}-\frac{\phi_0-\phi_{est}}{2\pi \tau_1} \\
    \delta f_2&=f_0-f_{est}-\frac{\phi_0-\phi_{est}}{2\pi \tau_2}.
\end{align*}

This system of two equations and two unknowns allows us to independently determine $\phi_0$ and $f_0$:
\begin{align*}
    \phi_0&=\phi_{est}+2\pi(\delta f_1-\delta f_2) \frac{\tau_1\tau_2}{\tau_1-\tau_2} \\
    f_0&=f_{est}+\frac{\delta f_1\tau_1 - \delta f_2\tau_2}{\tau_1-\tau_2}.
\end{align*}
The relevant parameter is $f_0$, which, when measured for transitions within the $S_{1/2}$ and $D_{5/2}$ manifolds, allows us to extract the $g$-factor ratio $g_{D52}/g_{S12}$.

To measure the ratio, we use the procedure described above to make repeated interleaved measurements of $f_{0,S}$ and $f_{0,D}$ for the transitions $S_{1/2} (m_J=-1/2\leftrightarrow +1/2)$ and $D_{5/2} (m_J=-5/2\leftrightarrow +5/2)$, respectively. The span of the $S_{1/2}$ manifold $f_{0,S}$ is measured by scanning $\delta f$ and interleaving experiment shots at Ramsey arms $\tau_1=100~\mu s$ and $\tau_2=1500~\mu s$. The measurement of $f_{0,S}$ uses a 729~nm pulse sequence consisting of two pulses (see blue transitions in Fig.~\ref{fig:levels}, right) with a total duration of $10~\mu s$ to create the desired superposition state. The span of the $D_{5/2}$ manifold $f_{0,D}$ is subsequently measured in the same manner but using a different 729~nm pulse sequence, consisting of four pulses with total duration of $32~\mu s$, for preparation of the desired superposition state (see red transitions in Fig.~\ref{fig:levels}, right). These sequential measurements of $f_{0,S}$ and $f_{0,D}$, each with a duration of $\sim 1$ minute, are repeated until the end of the measurement run. The $g$-factor ratio is then calculated as the mean of $g_{D52}/g_{S12} = f_{0,D}/(5f_{0,S})$ over the measurement run.

\begin{figure}
	\includegraphics[scale = 0.23]{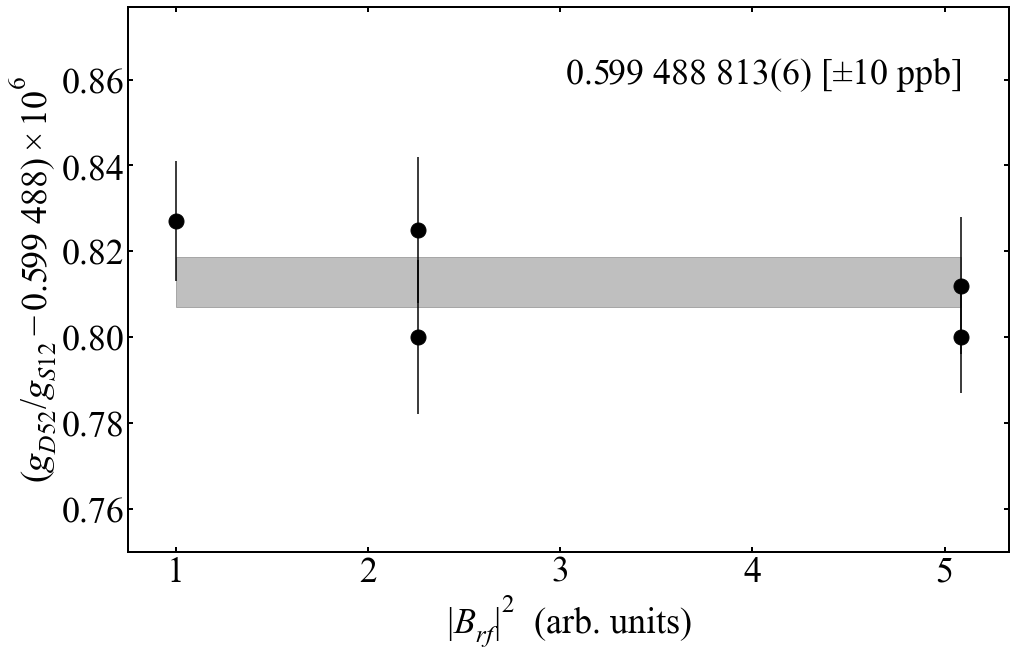}
	\caption{\label{fig:RFData} $g$-factor ratio measurements in the rf trap, taken at three trap rf amplitudes: $73~V_p$, $110~V_p$, and $165~V_p$. These rf amplitudes have an uncertainty of $10~V_p$ and are inferred from the measured radial mode frequencies, the rf frequency, and the dc potential. The data is plotted versus AC magnetic field amplitude $|B_{rf}|^2$ at the ion location. Each data point is the mean of a measurement run consisting of $>$~200 individual $g$-factor ratio measurements. The mean value 0.599~488~813(6) is the measured result for the rf trap.}
\end{figure}

The quantity $g_{D52}/g_{S12}$ is field-insensitive, so long as the individual measurements of $f_{0,S}$ and $f_{0,D}$ are made at the same magnetic field. Magnetic field drift between successive measurements of $f_{0,S}$ and $f_{0,D}$ can shift the measured value of $g_{D52}/g_{S12}$. We observe field drifts as large as $200$ pT/minute. We primarily attribute this drift to thermalization of the in-cryo permanent magnets after changing the trap rf amplitude, as discussed in the following paragraph. We cancel magnetic field drift to first order by averaging successive measurements of $f_{0,D}$, thereby extrapolating values of $f_{0,D}$ at the times corresponding to measurements of $f_{0,S}$. When computing $g_{D52}/g_{S12}$, we use the averaged values of $f_{0,D}$ and the measured values of $f_{0,S}$.

Certain systematic shifts in the measured value of $g_{D52}/g_{S12}$, such as the second-order Doppler shift, second-order Zeeman shift, and electric quadrupole shift, are canceled out due to our choice of measuring transitions between symmetric $m_J$ levels in each manifold. Potential significant systematic shifts (see Table \ref{table:error_budget}) include the AC Stark shift, due to leakage light at 397~nm, 729~nm, or 854~nm during the Ramsey delays, and the AC Zeeman shift, due to an oscillating magnetic field at the ion location originating from the trap rf~\cite{beloy_trap-induced_2023,joshi_characterization_2024,ivory_ac_2024}. Leakage light was measured using photon counters placed between the optical fiber and the vacuum window. Leakage was measured at $<1$~pW for 397~nm, 2~pW for 729~nm, and 1~pW for 854~nm. All leakage is detuned $>400$~MHz from any transition. This was determined to produce negligible AC Stark shifts compared to our measurement uncertainty (see Table~\ref{table:error_budget}).

To determine the magnitude of the AC Zeeman shift, we vary only the trap rf amplitude and measure $g_{D52}/g_{S12}$. The relative change in rf amplitude is determined by measuring the signal from a capacitive tap between the rf resonator and the ion trap. The data is shown in Fig.~\ref{fig:RFData}. We perform a linear fit to extract a slope corresponding to the AC Zeeman shift due to the trap rf. We find, however, that the slope is statistically insignificant. Thus we conclude that the AC Zeeman shift cannot be accurately determined due to the observed measurement uncertainty. We report the mean value 0.599~488~813(6) as the measured ratio $g_{D52}/g_{S12}$ for the rf trap. This result agrees with the higher-precision Penning trap result.

\textit{Conclusion.} In summary, we have measured the $g$-factor ratio between the $3^2D_{5/2}$ and $4^2S_{1/2}$ states of a single \Ca~ion in two distinct ion trap architectures. We find good agreement between the ratios measured in the two systems despite using distinct excitation techniques (microwave vs. optical) and a difference in trap magnetic field of more than three orders of magnitude. Our measurements also help to resolve the tension between the values of Ref.~\cite{chwalla_absolute_2009} and Refs.~\cite{ma_precision_2024,zhang_liquid-nitrogen-cooled_2026}, agreeing with the reported ratios in the latter publications. Combining our measured ratio with the most recent measurement of $g_{S12}$~\cite{tommaseo_mathsfg_scriptscriptstyle_2003} yields $g_{D52}=-1.200~330~46(5)$, where the $\sim45$~ppb fractional uncertainty of the $g_{S12}$ value dominates the total error of $g_{D52}$. The ratio reported here will provide a more stringent limit on the $3^2D_{5/2}$ $g$-factor following improved measurements of the $4^2 S_{1/2}$ $g$-factor. Beyond precision measurements, the metastable and ground state qubit excitation techniques demonstrated here may be advantageous for quantum information experiments that utilize diverse qubit encodings within a single ion species~\cite{allcock_omg_2021}.
 
The authors thank Nicholas Guise, Kyle Beloy, Ed Meyer, Bijaya Sahoo, and Murray Barrett for helpful discussions. Research is sponsored by the Laboratory Directed Research and Development Program of Oak Ridge National Laboratory, managed by UT-Battelle, LLC, for the US Department of Energy. Additionally, BJM and BCS acknowledge funding from the Office of Naval Research (Grant No. N00014-24-1-2360) and the Army Research Office (Grant No. W911NF-24-1-0355).

\bibliographystyle{apsrev4-2}
\bibliography{PenningRefs}

@article{britton_engineered_2012,
	title = {Engineered two-dimensional {Ising} interactions in a trapped-ion quantum simulator with hundreds of spins},
	volume = {484},
	issn = {0028-0836, 1476-4687},
	url = {http://www.nature.com/articles/nature10981},
	doi = {10.1038/nature10981},
	
	number = {7395},
	urldate = {2019-05-16},
	journal = {Nature},
	author = {Britton, Joseph W. and Sawyer, Brian C. and Keith, Adam C. and Wang, C.-C. Joseph and Freericks, James K. and Uys, Hermann and Biercuk, Michael J. and Bollinger, John J.},
	month = apr,
	year = {2012},
	pages = {489--492},
}

@article{koo_doppler_2004,
	title = {Doppler cooling of {Ca} + ions in a {Penning} trap},
	volume = {69},
	issn = {1050-2947, 1094-1622},
	url = {https://link.aps.org/doi/10.1103/PhysRevA.69.043402},
	doi = {10.1103/PhysRevA.69.043402},
	
	number = {4},
	urldate = {2019-05-16},
	journal = {Phys. Rev. A},
	author = {Koo, K. and Sudbery, J. and Segal, D. M. and Thompson, R. C.},
	month = apr,
	year = {2004},
	pages = {043402},
}

@article{crick_magnetically_2010,
	title = {Magnetically induced electron shelving in a trapped {Ca} + ion},
	volume = {81},
	issn = {1050-2947, 1094-1622},
	url = {https://link.aps.org/doi/10.1103/PhysRevA.81.052503},
	doi = {10.1103/PhysRevA.81.052503},
	
	number = {5},
	urldate = {2019-05-16},
	journal = {Phys. Rev. A},
	author = {Crick, D. R. and Donnellan, S. and Segal, D. M. and Thompson, R. C.},
	month = may,
	year = {2010},
	pages = {052503},
}

@article{shiga_diamagnetic_2011,
	title = {Diamagnetic correction to the 9 {Be} + ground-state hyperfine constant},
	volume = {84},
	issn = {1050-2947, 1094-1622},
	url = {https://link.aps.org/doi/10.1103/PhysRevA.84.012510},
	doi = {10.1103/PhysRevA.84.012510},
	
	number = {1},
	urldate = {2019-08-16},
	journal = {Phys. Rev. A},
	author = {Shiga, N. and Itano, W. M. and Bollinger, J. J.},
	month = jul,
	year = {2011},
	pages = {012510},
}

@article{tommaseo_mathsfg_scriptscriptstyle_2003,
	title = {The \${\textbackslash}mathsf\{g\_\{{\textbackslash}scriptscriptstyle {J}\}\}\$ -factor in the ground state of {Ca} \${\textasciicircum}{\textbackslash}mathsf\{+\}\$},
	volume = {25},
	issn = {1434-6060, 1434-6079},
	url = {http://www.springerlink.com/openurl.asp?genre=article&id=doi:10.1140/epjd/e2003-00096-6},
	doi = {10.1140/epjd/e2003-00096-6},
	
	number = {2},
	urldate = {2019-08-23},
	journal = {The European Physical Journal D - Atomic, Molecular and Optical Physics},
	author = {Tommaseo, G. and Pfeil, T. and Revalde, G. and Werth, G. and Indelicato, P. and Desclaux, J. P.},
	month = aug,
	year = {2003},
	pages = {113--121},
}

@article{beverini_g_1998,
	title = {g \_J {Factor} of neutral calcium {\textasciicircum}{3P} metastable levels},
	volume = {15},
	issn = {0740-3224, 1520-8540},
	url = {https://www.osapublishing.org/abstract.cfm?URI=josab-15-8-2206},
	doi = {10.1364/JOSAB.15.002206},
	
	number = {8},
	urldate = {2019-08-29},
	journal = {J. Opt. Soc. Am. B},
	author = {Beverini, N. and Maccioni, E. and Strumia, F.},
	month = aug,
	year = {1998},
	pages = {2206},
}

@article{mcmahon_doppler-cooled_2020,
	title = {Doppler-cooled ions in a compact reconfigurable {Penning} trap},
	volume = {101},
	issn = {2469-9926, 2469-9934},
	url = {https://link.aps.org/doi/10.1103/PhysRevA.101.013408},
	doi = {10.1103/PhysRevA.101.013408},
	
	number = {1},
	urldate = {2021-06-09},
	journal = {Phys. Rev. A},
	author = {McMahon, Brian J. and Volin, Curtis and Rellergert, Wade G. and Sawyer, Brian C.},
	month = jan,
	year = {2020},
	pages = {013408},
}

@article{brewer__2019,
	title = {Al + 27 {Quantum}-{Logic} {Clock} with a {Systematic} {Uncertainty} below 10 − 18},
	volume = {123},
	issn = {0031-9007, 1079-7114},
	url = {https://link.aps.org/doi/10.1103/PhysRevLett.123.033201},
	doi = {10.1103/PhysRevLett.123.033201},
	
	number = {3},
	urldate = {2021-09-08},
	journal = {Phys. Rev. Lett.},
	author = {Brewer, S. M. and Chen, J.-S. and Hankin, A. M. and Clements, E. R. and Chou, C. W. and Wineland, D. J. and Hume, D. B. and Leibrandt, D. R.},
	month = jul,
	year = {2019},
	pages = {033201},
}

@article{chwalla_absolute_2009,
	title = {Absolute {Frequency} {Measurement} of the {Ca} + 40 4 s {S} 1 / 2 2 − 3 d {D} 5 / 2 2 {Clock} {Transition}},
	volume = {102},
	issn = {0031-9007, 1079-7114},
	url = {https://link.aps.org/doi/10.1103/PhysRevLett.102.023002},
	doi = {10.1103/PhysRevLett.102.023002},
	
	number = {2},
	urldate = {2023-01-04},
	journal = {Phys. Rev. Lett.},
	author = {Chwalla, M. and Benhelm, J. and Kim, K. and Kirchmair, G. and Monz, T. and Riebe, M. and Schindler, P. and Villar, A. S. and Hänsel, W. and Roos, C. F. and Blatt, R. and Abgrall, M. and Santarelli, G. and Rovera, G. D. and Laurent, Ph.},
	month = jan,
	year = {2009},
	pages = {023002},
}

@article{allcock_omg_2021,
	title = {\textit{omg} blueprint for trapped ion quantum computing with metastable states},
	volume = {119},
	issn = {0003-6951, 1077-3118},
	url = {https://pubs.aip.org/apl/article/119/21/214002/1062039/omg-blueprint-for-trapped-ion-quantum-computing},
	doi = {10.1063/5.0069544},
	
	number = {21},
	urldate = {2024-01-03},
	journal = {Applied Physics Letters},
	author = {Allcock, D. T. C. and Campbell, W. C. and Chiaverini, J. and Chuang, I. L. and Hudson, E. R. and Moore, I. D. and Ransford, A. and Roman, C. and Sage, J. M. and Wineland, D. J.},
	month = nov,
	year = {2021},
	pages = {214002},
}

@article{pezze_quantum_2018,
	title = {Quantum metrology with nonclassical states of atomic ensembles},
	volume = {90},
	issn = {0034-6861, 1539-0756},
	url = {https://link.aps.org/doi/10.1103/RevModPhys.90.035005},
	doi = {10.1103/RevModPhys.90.035005},
	
	number = {3},
	urldate = {2024-01-09},
	journal = {Rev. Mod. Phys.},
	author = {Pezzè, Luca and Smerzi, Augusto and Oberthaler, Markus K. and Schmied, Roman and Treutlein, Philipp},
	month = sep,
	year = {2018},
	pages = {035005},
}

@article{arnold_prospects_2015,
	title = {Prospects for atomic clocks based on large ion crystals},
	volume = {92},
	issn = {1050-2947, 1094-1622},
	url = {https://link.aps.org/doi/10.1103/PhysRevA.92.032108},
	doi = {10.1103/PhysRevA.92.032108},
	
	number = {3},
	urldate = {2024-03-06},
	journal = {Phys. Rev. A},
	author = {Arnold, Kyle and Hajiyev, Elnur and Paez, Eduardo and Lee, Chern Hui and Barrett, M. D. and Bollinger, John},
	month = sep,
	year = {2015},
	pages = {032108},
}

@article{jain_penning_2024,
	title = {Penning micro-trap for quantum computing},
	issn = {0028-0836, 1476-4687},
	url = {https://www.nature.com/articles/s41586-024-07111-x},
	doi = {10.1038/s41586-024-07111-x},
	
	urldate = {2024-03-19},
	journal = {Nature},
	author = {Jain, Shreyans and Sägesser, Tobias and Hrmo, Pavel and Torkzaban, Celeste and Stadler, Martin and Oswald, Robin and Axline, Chris and Bautista-Salvador, Amado and Ospelkaus, Christian and Kienzler, Daniel and Home, Jonathan},
	month = mar,
	year = {2024},
}

@article{mcmahon_individual-ion_2024,
	title = {Individual-{Ion} {Addressing} and {Readout} in a {Penning} {Trap}},
	volume = {133},
	issn = {0031-9007, 1079-7114},
	url = {https://link.aps.org/doi/10.1103/PhysRevLett.133.173201},
	doi = {10.1103/PhysRevLett.133.173201},
	
	number = {17},
	urldate = {2024-11-15},
	journal = {Phys. Rev. Lett.},
	author = {McMahon, Brian J. and Brown, Kenton R. and Herold, Creston D. and Sawyer, Brian C.},
	month = oct,
	year = {2024},
	pages = {173201},
}

@article{cornell_mode_1990,
	title = {Mode coupling in a {Penning} trap: \textit{π} pulses and a classical avoided crossing},
	volume = {41},
	copyright = {http://link.aps.org/licenses/aps-default-license},
	issn = {1050-2947, 1094-1622},
	shorttitle = {Mode coupling in a {Penning} trap},
	url = {https://link.aps.org/doi/10.1103/PhysRevA.41.312},
	doi = {10.1103/PhysRevA.41.312},
	
	number = {1},
	urldate = {2025-03-18},
	journal = {Phys. Rev. A},
	author = {Cornell, Eric A. and Weisskoff, Robert M. and Boyce, Kevin R. and Pritchard, David E.},
	month = jan,
	year = {1990},
	pages = {312--315},
}

@article{zhiqiang_176_2023,
	title = {$^{\textrm{176}}$ {Lu}$^{\textrm{+}}$ clock comparison at the 10$^{\textrm{−18}}$ level via correlation spectroscopy},
	volume = {9},
	issn = {2375-2548},
	url = {https://www.science.org/doi/10.1126/sciadv.adg1971},
	doi = {10.1126/sciadv.adg1971},
	
	number = {18},
	urldate = {2026-01-21},
	journal = {Sci. Adv.},
	author = {Zhiqiang, Zhang and Arnold, Kyle J. and Kaewuam, Rattakorn and Barrett, Murray D.},
	month = may,
	year = {2023},
	pages = {eadg1971},
}

@article{moses_race-track_2023,
	title = {A {Race}-{Track} {Trapped}-{Ion} {Quantum} {Processor}},
	volume = {13},
	issn = {2160-3308},
	url = {https://link.aps.org/doi/10.1103/PhysRevX.13.041052},
	doi = {10.1103/PhysRevX.13.041052},
	
	number = {4},
	urldate = {2026-04-17},
	journal = {Phys. Rev. X},
	author = {Moses, S. A. and Baldwin, C. H. and Allman, M. S. and Ancona, R. and Ascarrunz, L. and Barnes, C. and Bartolotta, J. and Bjork, B. and Blanchard, P. and Bohn, M. and Bohnet, J. G. and Brown, N. C. and Burdick, N. Q. and Burton, W. C. and Campbell, S. L. and Campora, J. P. and Carron, C. and Chambers, J. and Chan, J. W. and Chen, Y. H. and Chernoguzov, A. and Chertkov, E. and Colina, J. and Curtis, J. P. and Daniel, R. and DeCross, M. and Deen, D. and Delaney, C. and Dreiling, J. M. and Ertsgaard, C. T. and Esposito, J. and Estey, B. and Fabrikant, M. and Figgatt, C. and Foltz, C. and Foss-Feig, M. and Francois, D. and Gaebler, J. P. and Gatterman, T. M. and Gilbreth, C. N. and Giles, J. and Glynn, E. and Hall, A. and Hankin, A. M. and Hansen, A. and Hayes, D. and Higashi, B. and Hoffman, I. M. and Horning, B. and Hout, J. J. and Jacobs, R. and Johansen, J. and Jones, L. and Karcz, J. and Klein, T. and Lauria, P. and Lee, P. and Liefer, D. and Lu, S. T. and Lucchetti, D. and Lytle, C. and Malm, A. and Matheny, M. and Mathewson, B. and Mayer, K. and Miller, D. B. and Mills, M. and Neyenhuis, B. and Nugent, L. and Olson, S. and Parks, J. and Price, G. N. and Price, Z. and Pugh, M. and Ransford, A. and Reed, A. P. and Roman, C. and Rowe, M. and Ryan-Anderson, C. and Sanders, S. and Sedlacek, J. and Shevchuk, P. and Siegfried, P. and Skripka, T. and Spaun, B. and Sprenkle, R. T. and Stutz, R. P. and Swallows, M. and Tobey, R. I. and Tran, A. and Tran, T. and Vogt, E. and Volin, C. and Walker, J. and Zolot, A. M. and Pino, J. M.},
	month = dec,
	year = {2023},
	pages = {041052},
}

@article{zhang_liquid-nitrogen-cooled_2026,
	title = {Liquid-{Nitrogen}-{Cooled} ​ 40 {Ca} + {Ion} {Optical} {Clock} with a {Systematic} {Uncertainty} of 4.4 × 10 − 19},
	volume = {136},
	issn = {0031-9007, 1079-7114},
	url = {https://link.aps.org/doi/10.1103/vngc-c1xv},
	doi = {10.1103/vngc-c1xv},
	
	number = {5},
	urldate = {2026-04-23},
	journal = {Phys. Rev. Lett.},
	author = {Zhang, Bao-lin and Ma, Zi-xiao and Huang, Yao and Han, Hui-li and Hu, Ru-ming and Wang, Yu-zhuo and Zhang, Hua-qing and Tang, Li-yan and Shi, Ting-yun and Guan, Hua and Gao, Ke-lin},
	month = feb,
	year = {2026},
	pages = {053202},
    note = {{S}upplementary {I}nformation},
}

@article{hartsell_design_2026,
	title = {Design and characterization of a cryogenic vacuum chamber for ion trapping experiments},
	volume = {128},
	issn = {0003-6951, 1077-3118},
	url = {https://pubs.aip.org/apl/article/128/3/034001/3377639/Design-and-characterization-of-a-cryogenic-vacuum},
	doi = {10.1063/5.0304948},
	
	number = {3},
	urldate = {2026-04-23},
	journal = {Applied Physics Letters},
	author = {Hartsell, D. M. and Gray, J. M. and Shappert, C. M. and Gostin, N. L. and McGill, R. A. and Tinkey, H. N. and Clark, C. R. and Brown, K. R.},
	month = jan,
	year = {2026},
	pages = {034001},
}

@article{blaum_high-accuracy_2006,
	title = {High-accuracy mass spectrometry with stored ions},
	volume = {425},
	copyright = {https://www.elsevier.com/tdm/userlicense/1.0/},
	issn = {03701573},
	url = {https://linkinghub.elsevier.com/retrieve/pii/S0370157305004643},
	doi = {10.1016/j.physrep.2005.10.011},
	
	number = {1},
	urldate = {2026-04-24},
	journal = {Physics Reports},
	author = {Blaum, Klaus},
	month = mar,
	year = {2006},
	pages = {1--78},
}

@misc{filzinger_multi-ion_2026,
	title = {A multi-ion optical clock with \${\textbackslash}mathbf\{5 {\textbackslash}times 10{\textasciicircum}\{-19\}\}\$ uncertainty},
	url = {http://arxiv.org/abs/2603.23446},
	doi = {10.48550/arXiv.2603.23446},
	
	urldate = {2026-04-24},
	publisher = {arXiv},
	author = {Filzinger, Melina and Steinel, Martin R. and Jiang, Jian and Bennett, Daniel and Mehlstäubler, Tanja E. and Peik, Ekkehard and Huntemann, Nils},
	month = mar,
	year = {2026},
	note = {arXiv:2603.23446 [physics]},
}

@article{noel_measurement-induced_2022,
	title = {Measurement-induced quantum phases realized in a trapped-ion quantum computer},
	volume = {18},
	issn = {1745-2473, 1745-2481},
	url = {https://www.nature.com/articles/s41567-022-01619-7},
	doi = {10.1038/s41567-022-01619-7},
	
	number = {7},
	urldate = {2026-04-24},
	journal = {Nat. Phys.},
	author = {Noel, Crystal and Niroula, Pradeep and Zhu, Daiwei and Risinger, Andrew and Egan, Laird and Biswas, Debopriyo and Cetina, Marko and Gorshkov, Alexey V. and Gullans, Michael J. and Huse, David A. and Monroe, Christopher},
	month = jul,
	year = {2022},
	pages = {760--764},
}

@misc{arnold_optical_2025,
	title = {Optical clocks with accuracy validated at the 19th digit},
	url = {http://arxiv.org/abs/2512.07346},
	doi = {10.48550/arXiv.2512.07346},
	
	urldate = {2026-04-24},
	publisher = {arXiv},
	author = {Arnold, K. J. and Lee, M. D. K. and Qi, Zhao and Qin, Qichen and Zhao, Zhang and Jayjong, N. and Barrett, M. D.},
	month = dec,
	year = {2025},
	note = {arXiv:2512.07346 [physics]},
}

@article{loschnauer_scalable_2025,
	title = {Scalable, {High}-{Fidelity} {All}-{Electronic} {Control} of {Trapped}-{Ion} {Qubits}},
	volume = {6},
	issn = {2691-3399},
	url = {https://link.aps.org/doi/10.1103/h4wk-v31j},
	doi = {10.1103/h4wk-v31j},
	
	number = {4},
	urldate = {2026-05-07},
	journal = {PRX Quantum},
	author = {Löschnauer, C.M. and Mosca Toba, J. and Hughes, A.C. and King, S.A. and Weber, M.A. and Srinivas, R. and Matt, R. and Nourshargh, R. and Allcock, D.T.C. and Ballance, C.J. and Matthiesen, C. and Malinowski, M. and Harty, T.P.},
	month = oct,
	year = {2025},
	pages = {040313},
}

@article{ma_precision_2024,
	title = {Precision determination of the oscillating-magnetic-field-induced second-order {Zeeman} shift of a single- {Ca} + 40 -ion optical clock},
	volume = {110},
	issn = {2469-9926, 2469-9934},
	url = {https://link.aps.org/doi/10.1103/PhysRevA.110.063102},
	doi = {10.1103/PhysRevA.110.063102},
	
	number = {6},
	urldate = {2026-05-12},
	journal = {Phys. Rev. A},
	author = {Ma, Zixiao and Zhang, Baolin and Huang, Yao and Hu, Ruming and Zeng, Mengyan and Gao, Kelin and Guan, Hua},
	month = dec,
	year = {2024},
	pages = {063102},
}

@article{beloy_trap-induced_2023,
	title = {Trap-{Induced} ac {Zeeman} {Shift} of the {Thorium}-229 {Nuclear} {Clock} {Frequency}},
	volume = {130},
	issn = {0031-9007, 1079-7114},
	url = {https://link.aps.org/doi/10.1103/PhysRevLett.130.103201},
	doi = {10.1103/PhysRevLett.130.103201},
	
	number = {10},
	urldate = {2026-05-15},
	journal = {Phys. Rev. Lett.},
	author = {Beloy, K.},
	month = mar,
	year = {2023},
	pages = {103201},
}

@article{joshi_characterization_2024,
	title = {Characterization of ion-trap-induced ac magnetic fields},
	volume = {110},
	issn = {2469-9926, 2469-9934},
	url = {https://link.aps.org/doi/10.1103/PhysRevA.110.063101},
	doi = {10.1103/PhysRevA.110.063101},
	
	number = {6},
	urldate = {2026-05-15},
	journal = {Phys. Rev. A},
	author = {Joshi, Manoj K. and Guevara-Bertsch, Milena and Kranzl, Florian and Blatt, Rainer and Roos, Christian F.},
	month = dec,
	year = {2024},
	pages = {063101},
}

@article{ivory_ac_2024,
	title = {{AC} {Zeeman} effect in microfabricated surface traps},
	volume = {95},
	issn = {0034-6748, 1089-7623},
	url = {https://pubs.aip.org/rsi/article/95/9/093202/3312284/AC-Zeeman-effect-in-microfabricated-surface-traps},
	doi = {10.1063/5.0204413},
	
	number = {9},
	urldate = {2026-05-15},
	journal = {Review of Scientific Instruments},
	author = {Ivory, M. and Nordquist, C. D. and Young, K. and Hogle, C. W. and Clark, S. M. and Revelle, M. C.},
	month = sep,
	year = {2024},
	pages = {093202},
}

@article{vogel_trap-assisted_2010,
	title = {Trap-assisted precision spectroscopy of forbidden transitions in highly-charged ions},
	volume = {490},
	copyright = {https://www.elsevier.com/tdm/userlicense/1.0/},
	issn = {03701573},
	url = {https://linkinghub.elsevier.com/retrieve/pii/S0370157309002919},
	doi = {10.1016/j.physrep.2009.12.007},
	
	number = {1-2},
	urldate = {2026-06-18},
	journal = {Physics Reports},
	author = {Vogel, Manuel and Quint, Wolfgang},
	month = may,
	year = {2010},
	pages = {1--47},
}

@article{mcmahon_efficient_2026,
	title = {Efficient three-dimensional sub-{Doppler} cooling of {Ca} + 40 in a {Penning} trap},
	volume = {114},
	issn = {2469-9926, 2469-9934},
	url = {https://link.aps.org/doi/10.1103/qq69-zdy3},
	doi = {10.1103/qq69-zdy3},
	
	number = {1},
	urldate = {2026-08-20},
	journal = {Phys. Rev. A},
	author = {McMahon, Brian J. and Sawyer, Brian C.},
	month = jul,
	year = {2026},
	pages = {013119},
}

@article{sahoo_relativistic_2017,
	title = {Relativistic coupled-cluster-theory analysis of unusually large correlation effects in the determination of g j factors in {Ca} +},
	volume = {96},
	copyright = {http://link.aps.org/licenses/aps-default-license},
	issn = {2469-9926, 2469-9934},
	url = {http://link.aps.org/doi/10.1103/PhysRevA.96.012511},
	doi = {10.1103/PhysRevA.96.012511},
	
	number = {1},
	urldate = {2026-08-20},
	journal = {Phys. Rev. A},
	author = {Sahoo, B. K. and Kumar, Pradeep},
	month = jul,
	year = {2017},
	pages = {012511},
}

@article{kumar_diverse_2018,
	title = {Diverse trends of electron correlation effects for properties with different radial and angular factors in an atomic system: a case study in {Ca}$^{\textrm{+}}$},
	volume = {51},
	issn = {0953-4075, 1361-6455},
	shorttitle = {Diverse trends of electron correlation effects for properties with different radial and angular factors in an atomic system},
	url = {https://iopscience.iop.org/article/10.1088/1361-6455/aaaa12},
	doi = {10.1088/1361-6455/aaaa12},
	
	number = {5},
	urldate = {2026-08-20},
	journal = {J. Phys. B: At. Mol. Opt. Phys.},
	author = {Kumar, Pradeep and Li, Cheng-Bin and Sahoo, B K},
	month = mar,
	year = {2018},
	pages = {055101},
}

@article{gan_oscillating-magnetic-field_2018,
	title = {Oscillating-magnetic-field effects in high-precision metrology},
	volume = {98},
	issn = {2469-9926, 2469-9934},
	url = {https://link.aps.org/doi/10.1103/PhysRevA.98.032514},
	doi = {10.1103/PhysRevA.98.032514},
	
	number = {3},
	urldate = {2026-08-20},
	journal = {Phys. Rev. A},
	author = {Gan, H. C. J. and Maslennikov, G. and Tseng, K.-W. and Tan, T. R. and Kaewuam, R. and Arnold, K. J. and Matsukevich, D. and Barrett, M. D.},
	month = sep,
	year = {2018},
	pages = {032514},
}

@article{arnold_precision_2020,
	title = {Precision {Measurements} of the {Ba} + 138 6 s {S} 2 1 / 2 − 5 d {D} 2 5 / 2 {Clock} {Transition}},
	volume = {124},
	issn = {0031-9007, 1079-7114},
	url = {https://link.aps.org/doi/10.1103/PhysRevLett.124.193001},
	doi = {10.1103/PhysRevLett.124.193001},
	
	number = {19},
	urldate = {2026-08-20},
	journal = {Phys. Rev. Lett.},
	author = {Arnold, K. J. and Kaewuam, R. and Chanu, S. R. and Tan, T. R. and Zhang, Zhiqiang and Barrett, M. D.},
	month = may,
	year = {2020},
	pages = {193001},
}

@article{monroe_programmable_2021,
	title = {Programmable quantum simulations of spin systems with trapped ions},
	volume = {93},
	issn = {0034-6861, 1539-0756},
	url = {https://link.aps.org/doi/10.1103/RevModPhys.93.025001},
	doi = {10.1103/RevModPhys.93.025001},
	
	number = {2},
	urldate = {2026-09-25},
	journal = {Rev. Mod. Phys.},
	author = {Monroe, C. and Campbell, W. C. and Duan, L.-M. and Gong, Z.-X. and Gorshkov, A. V. and Hess, P. W. and Islam, R. and Kim, K. and Linke, N. M. and Pagano, G. and Richerme, P. and Senko, C. and Yao, N. Y.},
	month = apr,
	year = {2021},
	pages = {025001},
}

@misc{guevara-bertsch_absolute_2026,
	title = {Absolute frequency measurement of the \${\textasciicircum}\{40\}\${Ca}\${\textasciicircum}\{+\}\$ clock transition using a {GNSS} link to the {SI} second},
	url = {http://arxiv.org/abs/2609.15844},
	doi = {10.48550/arXiv.2609.15844},
	
	urldate = {2026-09-25},
	publisher = {arXiv},
	author = {Guevara-Bertsch, M. and Joshi, M. K. and Hussain, M. I. and Blatt, R. and Roos, C. F.},
	month = sep,
	year = {2026},
	note = {arXiv:2609.15844 [physics.atom-ph]},
}

\end{document}